\documentclass[useAMS,usenatbib,fleqn]{mnras}

\usepackage[T1]{fontenc}
\usepackage{ae,aecompl}

\usepackage{graphicx}	
\usepackage{amssymb}	

\usepackage{times}
\usepackage{epstopdf}
\usepackage{subfigure}
\usepackage{multirow}
\usepackage{pdflscape}
\usepackage{rotating}

\usepackage{xcolor}

\usepackage{anyfontsize}

\usepackage{scalerel}
\usepackage{tikz}
\usetikzlibrary{svg.path}

\usepackage{newtxtext,newtxmath}

\newcommand{\orcidpng}[1]{\href{https://orcid.org/#15}{\includegraphics[scale=0.1]{/Users/maurosereno/Documents/bozze/ORCID-iD_icon-128x128.png}}}

\definecolor{orcidlogocol}{HTML}{A6CE39}
\tikzset{
  orcidlogo/.pic={
    \fill[orcidlogocol] svg{M256,128c0,70.7-57.3,128-128,128C57.3,256,0,198.7,0,128C0,57.3,57.3,0,128,0C198.7,0,256,57.3,256,128z};
    \fill[white] svg{M86.3,186.2H70.9V79.1h15.4v48.4V186.2z}
                 svg{M108.9,79.1h41.6c39.6,0,57,28.3,57,53.6c0,27.5-21.5,53.6-56.8,53.6h-41.8V79.1z M124.3,172.4h24.5c34.9,0,42.9-26.5,42.9-39.7c0-21.5-13.7-39.7-43.7-39.7h-23.7V172.4z}
                 svg{M88.7,56.8c0,5.5-4.5,10.1-10.1,10.1c-5.6,0-10.1-4.6-10.1-10.1c0-5.6,4.5-10.1,10.1-10.1C84.2,46.7,88.7,51.3,88.7,56.8z};
  }
}

\newcommand\orcid[1]{\href{https://orcid.org/#1}{\mbox{\scalerel*{
\begin{tikzpicture}[yscale=-1,transform shape]
\pic{orcidlogo};
\end{tikzpicture}
}{|}}}}

\newcommand{\beq}{\begin{equation}}
\newcommand{\eeq}{\end{equation}}

\def\gs{\mathrel{\lower0.6ex\hbox{$\buildrel {\textstyle >}\over{\scriptstyle \sim}$}}}
\def\ls{\mathrel{\lower0.6ex\hbox{$\buildrel {\textstyle <}\over{\scriptstyle \sim}$}}}
\newcommand{\simgt}{\lower.5ex\hbox{$\; \buildrel > \over \sim \;$}}
\newcommand{\simlt}{\lower.5ex\hbox{$\; \buildrel < \over \sim \;$}}

\DeclareRobustCommand{\VAN}[3]{#2}
\let\VANthebibliography\thebibliography
\def\thebibliography{\DeclareRobustCommand{\VAN}[3]{##3}\VANthebibliography}

\title[Thermalisation of galaxy clusters]{The thermalisation of massive galaxy clusters}

\author[M. Sereno et al.]{
Mauro Sereno \orcid{0000-0003-0302-0325},$^{1,2}$\thanks{E-mail: mauro.sereno@inaf.it} 
Lorenzo Lovisari \orcid{0000-0002-3754-2415},$^{1,3}$
Weiguang Cui \orcid{0000-0002-2113-4863},$^{4}$
and Gerrit Schellenberger \orcid{0000-0002-4962-0740}$^{3}$\\
$^1$INAF - Osservatorio di Astrofisica e Scienza dello Spazio di Bologna, via Piero Gobetti 93/3, I-40129 Bologna, Italy\\
$^2$INFN - Sezione di Bologna, viale Berti Pichat 6/2, I-40127 Bologna, Italy\\
$^3$Center for Astrophysics $|$ Harvard $\&$ Smithsonian, 60 Garden Street, Cambridge, MA 02138, USA\\
$^4$Institute for Astronomy, University of Edinburgh, Royal Observatory, Edinburgh EH9 3HJ, United Kingdom
}

\date{Accepted 2021 August 18. Received 2021 August 2; in original form 2021 June 22}

\begin{document}
\label{firstpage}
\pagerange{\pageref{firstpage}--\pageref{lastpage}}
\maketitle

\begin{abstract}
In the hierarchical scenario of structure formation, galaxy clusters are the ultimate virialised products in mass and time. 
Hot baryons in the intracluster medium (ICM) and cold baryons in galaxies inhabit a dark matter dominated halo. Internal processes, accretion, and mergers can perturb the equilibrium, which is established only at later times. 
However, the cosmic time when thermalisation is effective is still to be assessed. 
Here we show that massive clusters in the observed universe attained an advanced thermal equilibrium $\sim~1.8~\text{Gyr}$ ago, at redshift $z =0.14\pm0.06$, when the universe was $11.7\pm0.7~\text{Gyr}$ old.  Hot gas is mostly thermalised after the time when cosmic densities of matter and dark energy match.
We find in a statistically nearly complete and homogeneous sample of 120 clusters from the {\it Planck} Early Sunyaev-Zel'dovich (ESZ) sample that the kinetic energy traced by the galaxy velocity dispersion is a faithful probe of the gravitational energy since a look back time of at least $\sim5.4~\text{Gyr}$, whereas the efficiency of hot gas in converting kinetic to thermal energy, as measured through X-ray observations in the core-excised area within $r_{500}$, steadily increases with time. The evolution is detected at the $\sim 98$ per cent probability level.
Our results demonstrate that halo mass accretion history plays a larger role for cluster thermal equilibrium than radiative physics. The evolution of hot gas is strictly connected to the cosmic structure formation. 
\end{abstract}





\section{Introduction}

In the $\Lambda$CDM (Cold Dark Matter with a cosmological constant $\Lambda$) scenario of structure formation, small objects are the first to decouple, collapse, and virialise \citep{voi05}. They later gather into larger haloes. Galaxy clusters are the end points of this process and the larger halos to near virial equilibrium. 

The growth of clusters directly traces the process of structure formation in the universe. Perturbation growth, when positive density perturbations exceed the mean density, is more rapid in the matter dominated universe. Growth occurs mostly before the time when dark matter density equals dark energy at $z\sim0.33$ (corresponding to a look-back time $t_\text{lb}\sim3.7~\text{Gyr}$), after which the universe enters its phase of accelerated expansion and the perturbation growth slows down.

Cluster galaxies (which contributes to $\sim 3$ per cent of the cluster total mass) and gas ($\sim 12$ per cent) settle in the gravitational potential mainly shaped by the DM halo ($\sim 85$ per cent) \citep{voi05}. In a self-similar scenario of formation and evolution, galaxy clusters are unchanging in form and appearance if rescaled \citep{ber85,kai86}. As the result of gravitationally driven processes, the gas thermalises due to shock waves  and reaches hydrostatic equilibrium when gravity is balanced by a pressure-gradient force produced by thermal motion. In isolated systems, the galaxies move in near virial equilibrium, with twice their average total kinetic energy equaling the absolute value of the average total potential energy.

Gravitational interaction and matter accretion can perturb clusters. Major mergers disrupt the virial equilibrium. The ICM takes more time to adjust than galaxies or collisionless DM \citep{clo+al04,ok+um08}. Baryonic processes can impact the energy budget \citep{sta+al10,tru+al18}. Radiative emission makes the gas cooler. Star formation removes cold, low entropy gas. Feedback from AGN (Active Galactic Nucleus) and stellar winds can heat the surrounding ICM and push hotter gas to larger radii. Self-similarity is altered in the inner, cool cores, which do not change in size or density with time \citep{mcd+al17}.

The epoch of thermalisation, when hot gas reaches the thermal equilibrium in the DM potential, is then a combined result of, on one side, cosmic structure evolution and formation, and, on the other side, local gravitational interactions, and baryonic processes. Here, we analyse the energy content in a representative sample of the most massive haloes in the observable universe derived from the {\it Planck} Early Sunyaev-Zel'dovich (ESZ) clusters, a statistically complete and homogeneous sample of 188 confirmed clusters from the first ten months of the {\it Planck} survey \citep{planck_early_VIII}. Samples with a well defined selection function are crucial in our understanding of structure formation. SZ selected galaxy clusters are viewed as reliable tracers of the cosmological halo mass function and preserve the diversity of the full halo population \citep{ros+al16}. The SZ signal is correlated to the mass, and it is not significantly affected by dynamical state or mergers \citep{kra+al06}. The SZ brightness is independent of the distance and clusters can be found up to high redshifts \citep{planck_2015_XXVII,ble+al15}. SZ clusters at intermediate redshifts are then approximatively mass-selected. The halo properties of {\it Planck} clusters are found to be in agreement with the $\Lambda$CDM model  \citep{ser+al17_psz2lens}. 



We measure the energy budget within $r_{500}$, a region that covers $\sim 60$ per cent of the viral radius and encompasses $\sim 50$-$60$ per cent of the virial mass. This region is very sensitive to the cluster formation and evolution. The radius $r_{500}$ is nearly twice the scale radius within which the mass density remains approximately constant during the slow accretion stage which follows an early fast accretion phase \citep{go+re75,wec+al02,lud+al13,mos+al19}. Thermodynamic properties within $r_{500}$ are remarkably self-similar, as found in a sample of local, SZ selected, massive clusters \citep{ghi+al19}. The intrinsic scatter of thermodynamic quantities is minimised in the $0.2 \lesssim r/r_{500} \lesssim 0.8$ range, the region where the gas is mostly virialised and baryonic effects are negligible. In the inner region ($r/r_{500} \lesssim 0.3$), baryonic effects, e.g. cooling or AGN feedback, lead to a substantial scatter within the cluster population, while in the outer regions beyond $r_{500}$ the scatter is mostly driven by different accretion rates. The core excised temperature within $r_{500}$ can then be considered a reliable probe of the gas thermal energy.



Theoretical predictions are naturally expressed in terms of energies \citep{gio+al13}, but most of the focus in recent years has been on mass. Paradoxically, the cluster mass is more difficult to measure than the potential energy, which is more directly related to the observables  \citep{tch+al20}. Here, we recast usual mass--temperature--velocity dispersion relations in terms of potential--thermal-kinetic energy, which simplifies the study of the time evolution.

As reference cosmological framework, we assume the concordance flat $\Lambda$CDM model with total matter density parameter $\Omega_\text{M0}=1-\Omega_{\Lambda 0}=0.3$, and Hubble constant $H_0=70~\text{km~s}^{-1}\text{Mpc}^{-1}$. The present-day age of the Universe in this model is $\sim13.5~\text{Gyr}$. A generic quantity $O_{\Delta}$ is computed within $r_{\Delta}$, the radius within which the average cluster matter density is $\Delta$ times the critical density of the universe at the cluster redshift, $\rho_\mathrm{cr}=3H(z)^2/(8\pi G)$, where $H(z) (= H_0 E_z)$ is the redshift dependent Hubble parameter.


\section{Model} 
\label{sec_mode}

We consider the potential, $E_\phi$, kinetic, $E_\text{k}$, and thermal, $E_\text{th}$, energy of a test particle with mean molecular weight $\mu \sim 0.59$. We evaluate energies as proxies which do not require any assumption on density profile, symmetry, isothermality, or equilibrium. This approach is suitable for a diverse cluster population. The proxies are expressed in terms of integrated or weighted cluster properties: the mass $M_{500}$ and the radius $r_{500}$ for  $E_\phi$; the weighted core-excised gas temperature for $E_\text{th}$; the projected galaxy velocity dispersion for $E_\text{k}$. As energy proxies for the typical energies of a test particle within $r_{500}$, we consider
\begin{eqnarray}
E_\phi & = &	-\frac{3}{2} G  \frac{\mu m_\text{p} M_{500}}{r_{500}}	 \,,  \label{eq_en_phi}\\
E_\text{th} & = & \frac{3}{2} k_\text{B} T_\text{g} \,,  \label{eq_en_th} \\
E_\text{k} & = & \frac{3}{2} \mu m_\text{p} \sigma_\text{1D}^2  \,,  \label{eq_en_k}
\end{eqnarray}
where $G$ is the gravitational constant, $k_\text{B}$ is the Boltzmann constant, and $m_\text{p}$ the proton mass. The proxy for the potential energy is given in terms of the circular velocity and it is directly related to the DM dominated potential. The choice of the numerical factor in Eq.~(\ref{eq_en_phi}) is arbitrary but it does not affect our main conclusions, which are based on relative evolution and not on absolute calibration. The thermal energy is expressed in terms of the weighted temperature of the hot gas. The proxy $E_\text{k}$ tracks the kinetic energy of the cold baryons in galaxies. $\sigma_\text{1D}$ is the line of sight galaxy velocity dispersion.

Cluster evolution is more conveniently expressed in terms of potential energy than mass. The relations between energies in the self-similar model are redshift independent. Furthermore, the gravitational potential is more directly connected to observations \citep{tch+al20}, and measurements are more precise (the relative statistical uncertainty on the potential is nearly two thirds of the mass uncertainty). The connection between the thermal--potential energy and the mass--temperature relations is detailed in App.~\ref{sec_m-t}.

Different scenarios of cluster formation and evolution can be summarised thanks to the energy proxies in Eqs.~(\ref{eq_en_phi}, \ref{eq_en_th}, \ref{eq_en_k}). For self-similar, steady, and isolated clusters in equilibrium, the halo is isothermal with isotropic velocity dispersion. The thermal energy is expected to equal the kinetic energy and it is one half of the absolute value of the potential energy. For relaxed haloes, these relations hold at any time and there is no redshift evolution. Relations between energy proxies can be derived for self-similar clusters in virial and hydrostatic equilibrium \citep{voi05}. For the singular isothermal sphere (SIS),
\begin{equation}
\label{eq_eq_vir}
E_\text{th} = E_\text{k}= -\frac{E_\phi}{2} \ .
\end{equation}

However, clusters are neither isolated, nor steady, nor isothermal. They reside in the nodes of the cosmic web, and continuously grow through accretion of matter along filaments from outer environment or mergers. Terms originating from the flux of inertia through the boundary and the pressure at the boundary enter the virial equilibrium \citep{voi05,evr+al08,fuj+al18a}. 

Theoretical predictions for scaling relations between energies can be formulated for interacting clusters. For clusters continuously accreting mass from the surrounding environment, the relation between thermal and potential energy reflects the central structure \citep{fuj+al18a}. In the inside-out halo growth scenario, the intracluster medium is heated up in the fast-rate growth phase when the shape of the potential well is established, and the dark-matter halo structure and the gas preserve the memory of the formation time \citep{fuj+al18a}. Based on a self-similar model for secondary infall and accretion onto an initially overdense perturbation \citep{ber85}, the mass-temperature relation of accreting clusters can be predicted in terms of the fundamental plane and the mass dependence of the halo concentration \citep{fuj+al18b}. Even though this model does not assume virial equilibrium, the thermal and the potential energy are still approximately proportional, 
\beq
E_\text{th} \propto E_\phi \,
\eeq 
where the temperature is mass-weighted within the central region. The solution is nearly time-independent and close to the relation for isolated systems, see Eq.~(\ref{eq_eq_vir}), with some marginal evidence for positive evolution \citep{fuj+al18b}.


Baryonic processes and accretion from the surrounding large-scale structure impact the relation between kinetic and potential energy. Galaxies are biased tracers of the DM particles. Since we are interested in the equilibrium status of the cold baryons, we express the kinetic energy in terms of the velocity dispersion of galaxies rather than DM particles. The relation between velocity dispersion and mass for not isolated halos has been addressed through numerical simulations \citep{evr+al08,mun+al13,sar+al13}. Pressure terms enter the virial relation. Dynamical friction, merging with the central galaxy, tidal disruption, and accretion from the surrounding large-scale structure impact the relation between galaxy kinetic energy and dark matter potential. The analysis of the relation between the mass of galaxy clusters and the velocity dispersion in hydrodynamical or semi-analytical simulations shows some evidence for a mass dependent velocity dispersion bias, but the bias is found to be time independent. The evolution in redshift is remarkably self-similar \citep{mun+al13,sar+al13}, whether the tracers are DM particles, subhaloes, or galaxies \citep{mun+al13}, and independently of the adopted gas physics and feedback modelling \citep{mun+al13}. As a result, the relation between kinetic and potential energy (in arbitrary units) is slightly steeper than the self-similar prediction at the low mass end  \citep{mun+al13}, 
\beq
\label{eq_en_phi_en_k}
E_\text{k} \sim | E_\phi |^{1.09} \ , 
\eeq
with no evidence for redshift evolution in the $0<z<0.6$ range. 



\section{Data}
\label{sec_data}

\subsection{Sample}

The ESZ-XMM clusters, see Fig.~\ref{fig_ESZ_sample}, are a representative subsample in mass and redshift of 120 ESZ clusters selected to ensure that $r_{500}$ is completely covered by {\it XMM-Newton} observations \citep{lov+al17,lov+al20}. 

The ESZ-XMM clusters are morphologically classified based on X-ray concentration and centroid shift within $r_{500}$ \citep{lov+al17}. Regular clusters in their X-ray features are not necessarily dynamically relaxed in the virial region \citep{men+al14}. As the regular subsample, we consider one-third of the full sample including the more regular ESZ-XMM clusters. This fraction is comparable to the expected fraction of relaxed halos \citep{cui+al18}. In the following, the remaining two-thirds are considered irregular. The subsample distribution follows the parental ones ($\sim 99$ and $47$ per cent in mass and redshift for the regular clusters according to the Kolmogorov-Smirnov test), with a slight overabundance of regular clusters at low redshifts. 

The mass and redshift distributions of the ESZ-XMM systems are representative of the whole ESZ sample of 188 galaxy clusters \citep{lov+al17}, even though the requirement of complete X-ray coverage can excise very near or massive clusters which cover a very large sky area (e.g. the Coma cluster). The minimum redshift of ESZ-XMM is $z\sim 0.06$ and most of the not included ESZ clusters are at $z\la 0.1$.

ESZ-XMM comprises some of the most massive clusters in the {\it Planck}-observed universe at any epoch since $z=0.55$. This covers the last 5.4~Gyr of cosmic history and encompasses the dark matter - dark energy equality at $z\sim 0.33$.

\subsection{X-rays}

Accurate spectroscopic core-excised temperatures of the gas, $T_\text{g}$, enable us to measure the thermal energy. Temperatures in the $[0.15-1]~r_{500}$ range were determined by integrating the deprojected temperature profiles along the line of sight weighted by the emission measure and accounting for the detector response \citep{lov+al20}. These temperatures are in good agreement with estimates based on a single spectral extraction. 
The gravitational well is estimated thanks to X-ray halo masses, $M_{500}$, obtained by assuming hydrostatic equilibrium and a parametric model for the density profile \citep{lov+al20}.

\subsection{Velocity dispersions} 

For the estimation of the kinetic energy, we rely on galaxy velocity dispersions, $\sigma_\text{1D}$, retrieved from the meta-catalogue SC--Sigma Catalog \citep{se+et15_comalit_IV}, where we find 67 ESZ-XMM clusters\footnote{We used the \textsc{SC-single\_v2.0.dat} version. Updates are periodically made available at  \url{http://pico.oabo.inaf.it/\textasciitilde sereno/CoMaLit/sigma/}.}.
The median number of confirmed spectroscopic members per cluster is 75. The clusters with measured $\sigma_\text{1D}$  make an unbiased subsample of ESZ-XMM. According to the Kolmogorov-Smirnov test, their mass and redshift distribution is compatible with that of the entire sample with a probability of $\sim90$ or 87 per cent, respectively.


\begin{figure}
\resizebox{\hsize}{!}{\includegraphics{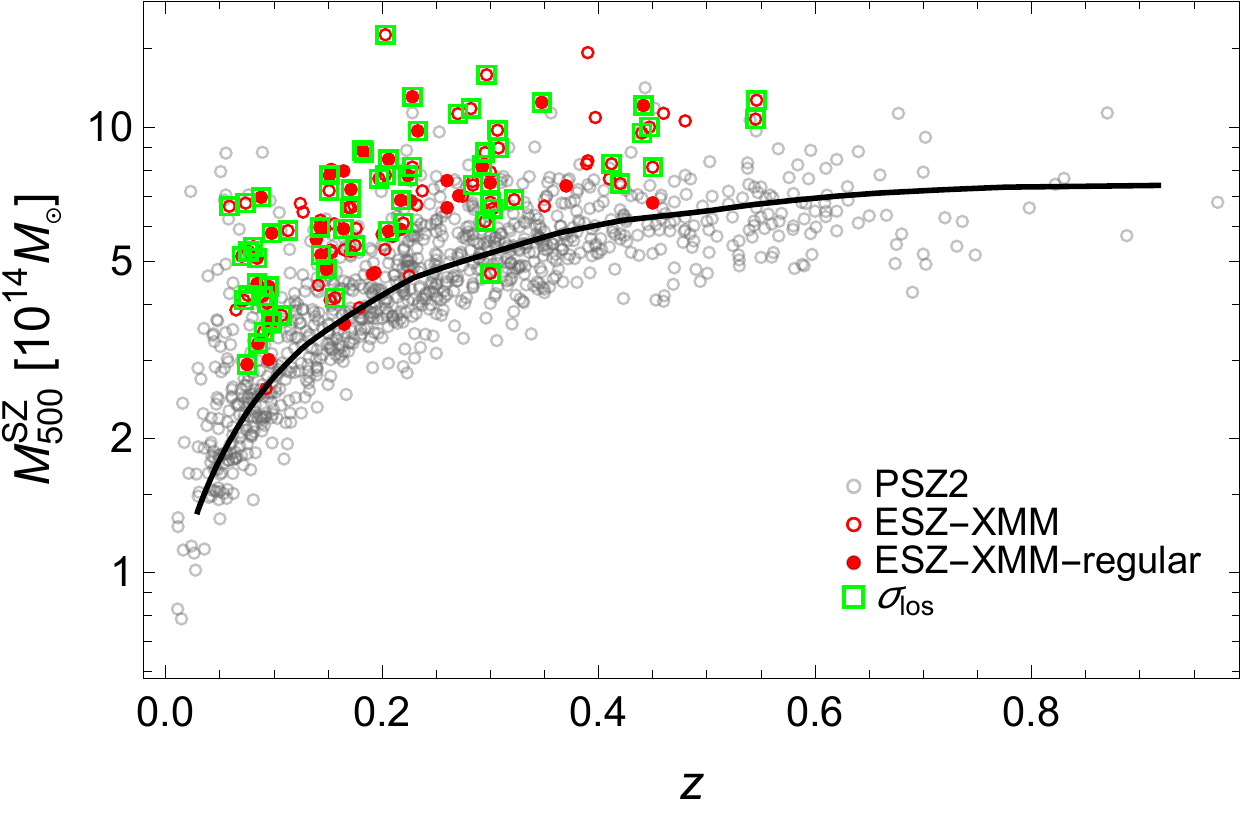}}
\caption{
Sample. Shown is the distribution in the $M^\text{SZ}_{500}$--$z$ plane of the ESZ-XMM clusters (red circles). The X-ray morphologically regular clusters are marked as red disks. The 1094 PSZ2 clusters with known-redshift counterparts
are shown as grey circles. The ESZ-XMM clusters with measured galaxy velocity dispersion are marked with a green square. The solid line indicates the average 80 per cent survey completeness limit for PSZ2. The mass $M^\text{SZ}_{500}$ is based on a SZ proxy. For the five ESZ-XMM clusters with no counterpart in the PSZ2 catalog, we plot the hydrostatic mass. $M^\text{SZ}_{500}$ is computed within $r_{500}$, the radius within which the average cluster matter density is $500$ times the critical density of the universe at the cluster redshift.
}
\label{fig_ESZ_sample}
\end{figure}

\subsection{Simulations} 

We compare our sample to simulated clusters from the Three Hundred project \citep[the 300,][]{cui+al18,mos+al19}. The project models large galaxy clusters and their environment with full-physics hydrodynamical re-simulations. The smoothed particle hydrodynamics scheme includes an artificial conduction accounting for gas-dynamical instabilities and mixing processes. Gas cooling and heating, star formation, stellar population properties and chemistry, stellar feedback, supermassive black hole growth, and AGN feedback are modelled with suitable prescriptions. From the simulated sample, we select 720 haloes by matching to each ESZ-XMM cluster the 6 haloes closer in redshift and mass, so that the extracted simulated clusters follow the observed distribution.


\section{Regression} 
\label{sec_regr}

Multi-wavelength investigations can efficiently constrain cluster properties \citep{man+al15,xxl_XXXVIII_ser+al19,far+al19}. We retrieve the evolution of the energy budget through the joint analysis of the scaling relations between the potential, thermal, and kinetic energy, $E_\phi$--$E_\text{th}$--$E_\text{k}$. We employ a Bayesian inference method which accounts for correlated intrinsic scatters, selection effects, and parent cluster distribution \citep{se+et15_comalit_IV,ser16_lira,ser+al20_hscxxl}. Kinetic energies for clusters without measured velocity dispersion are treated as latent variables \citep{se+et17_comalit_V}, which preserves the statistical completeness and homogeneity of the sample. Even though the SZ signal is proportional to the thermal energy content of the ICM, the experimental process which detected the ESZ clusters is independent from the XMM measurements and from the estimates of the velocity dispersion, which makes the Malmquist bias negligible.

Scaling relations between energies are modelled as power laws. The redshift dependence is expressed in terms of the cosmic critical density through the Hubble parameter. This is a standard form to model deviations from self-similarity. In the redshift range $0<z<0.6$, the growth factor of linear perturbations, $D_1$, or the expansion factor, $a=1/(1+z)$, can be conveniently approximated as power-laws of the Hubble parameter \citep{se+et15_comalit_IV}, which then provides a flexible parameterisation\footnote{If needed to compare with previous results, we consider that $E_z \sim (1+z)^{0.75}$ in the redshift range $0\la z\la 1$.}.

For each cluster, we consider the known correlated measurement uncertainties and the unknown correlated intrinsic scatters. Uncertainties on spectroscopic redshifts are treated as negligible.
 
We analyse the scaling relations, the distribution of cluster energies, $p (|E_\phi|, z)$, and the intrinsic scatter covariance matrix, $\Sigma_{E_\text{th},E_\text{k},E_\phi}$ \citep{ser+al20_hscxxl}. The relations and quantities to be fitted are
\begin{eqnarray}
\log \frac{E_\text{th}}{E_\text{p}}	& = & \alpha_{E_\text{th} | E_\phi} +  \beta_{E_\text{th} | E_\phi} \log \frac{|E_\phi|}{E_\text{p}} +  \gamma_{E_\text{th} | E_\phi} \log \frac{H_z}{H_\text{p}} \, ,\\
\log \frac{E_\text{k}}{E_\text{p}}		& = & \alpha_{E_\text{k} | E_\phi} +  \beta_{E_\text{k} | E_\phi} \log \frac{|E_\phi|}{E_\text{p}} +  \gamma_{E_\text{k} | E_\phi} \log \frac{H_z}{H_\text{p}} \, ,
\end{eqnarray}
and
\begin{eqnarray}
\mathbf{\Sigma}_{E_\text{th},E_\text{k},E_\phi} & & \\
p \left(\log \frac{|E_\phi|}{E_\text{p}}, z \right)  & =  & {\cal N} (\mu(z), \sigma(z)) \, ,  \label{eq_fit_dist}
\end{eqnarray}
where $\log$ is the logarithm base 10, $E_\text{p}=3\times 10^{-15}~\text{J}(\sim 18.7~\text{keV})$, and $H_\text{p}=H(z_\text{p}=0.2)$. 

SZ selected samples are not strictly mass selected. The mass limit of ESZ is scattered and redshift dependent, see Fig.~\ref{fig_ESZ_sample}. To properly deal with the sample selection effects and disentangle mass from time evolution, we fit the parameters of the covariate distribution, i.e. the potential energy distribution, at the same time of the scaling relations, see Eq.~(\ref{eq_fit_dist}). The energy distribution is modelled as a time evolving Gaussian distribution, ${\cal N}$, with redshift dependent mean, $\mu$, and dispersion, $\sigma$,
\begin{eqnarray}
\mu (z)      &  =  & \bar{\mu} + \gamma_{\mu,D}  \left[ D_\mathrm{A}(z)/D_\mathrm{A}(z_\text{p}) \right]  \, , \label{eq_muz}\\
\sigma (z) & =  &  \sigma_\text{p}  \left[ D_\mathrm{A}(z)/D_\mathrm{A}(z_\text{p}) \right]^{\gamma_{\sigma,D}} \, , \label{eq_siz}
\end{eqnarray}
where $D_\mathrm{A}$ is the angular diameter distance. 

We consider non-informative priors \citep{xxl_XXXVIII_ser+al19,ser+al20_hscxxl}: uniform distributions for the normalisations $\alpha$ and Gaussian mean $\bar{\mu}$; the Student's $t_1$ distribution with one degree of freedom for the slopes $\beta$ and $\gamma$; the Gamma distribution for the inverse of the variance, $\sigma_\text{p} ^{-2}$; the scaled Wishart distribution with two degrees of freedom for the inverse of the intrinsic scatter covariance matrix. 

In total, we fit 16 parameters to 120 redshifts, potential and thermal energies, and 67 kinetic energies. 

When not stated otherwise, central values and dispersions are computed as bi-weight estimators of the parameter marginalised posteriori distribution \citep{bee+al90}. Probabilities are computed considering the marginalised posteriori distributions.


\begin{figure}
\begin{tabular}{c}
\resizebox{\hsize}{!}{\includegraphics{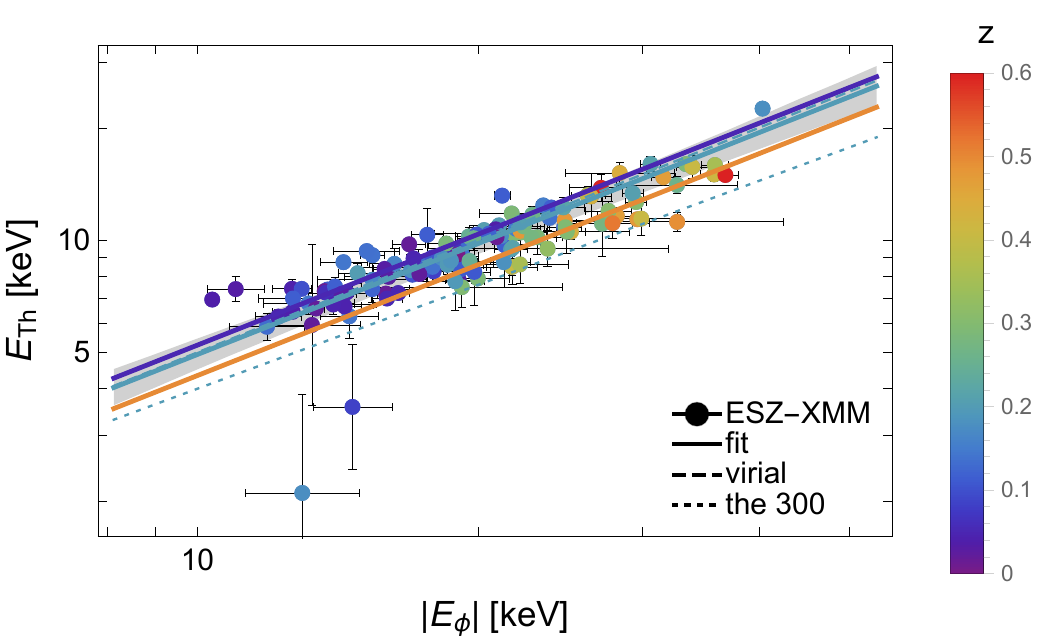}}\\
\resizebox{\hsize}{!}{\includegraphics{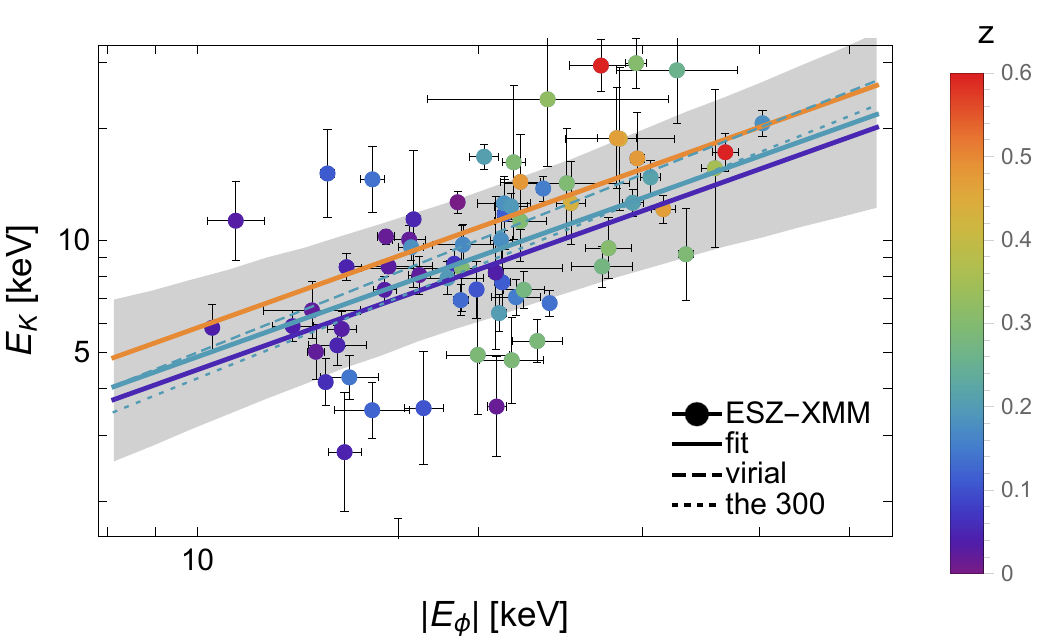}} 
\end{tabular}
\caption{Scaling. Shown is the scaling of the (absolute value of the) potential energy $|E_\phi|$ with the thermal energy $E_\text{th}$ (upper panel), or the kinetic energy $E_\text{k}$ (lower panel). The solid, dashed, or dotted lines show the fit to the ESZ-XMM sample, the virial expectation, and the fit to the 300 simulations, respectively. Points and lines are colour-coded according to the redshift (bar legend on the right side). The blue, cyan, or orange lines are plotted at $z=0.05$, 0.2, and 0.5, respectively. The shadowed region encompasses the 68.3 per cent confidence region around the $z=0.2$ fit.
Lower redshift clusters (bluer points) are systematically on the left side of the higher-$z$ haloes (redder points) in the $E_\text{th}$--$|E_\phi|$ plane, denoting hotter clusters at low redshift.
}
\label{fig_E_th_e_k_E_pot}
\end{figure}

\begin{table}
\caption{Scaling relations. The scaling parameters are derived by fitting the expression $E_i/E_\text{p} = 10^\alpha |E_\phi/E_\text{p} |^\beta [H(z)/H(z_\text{p})]^\gamma$, where $E_i$ is 
$E_\text{th}$ or $E_\text{k}$, $E_\text{p}=3\times 10^{-15}~\text{J}(\sim 18.7~\text{keV})$, $z_\text{p}=0.2$, and $H(z)$ is the redshift dependent Hubble parameter.}
\label{tab_scaling}
\centering
{\footnotesize
\begin{tabular}[c]{l  r@{$\,\pm\,$}l    r@{$\,\pm\,$}l  r@{$\,\pm\,$}l }
\hline
	\noalign{\smallskip}  
   &  \multicolumn{2}{c}{$10^\alpha$}	&  \multicolumn{2}{c}{$\beta$}	&  \multicolumn{2}{c}{$\gamma$}  \\
 	$E_\text{th}$-$E_\phi$ & \multicolumn{6}{c}{}	\\
    	 \hline
	\noalign{\smallskip}      
ESZ-XMM         	&	0.50        	&	0.01                	&	0.99	&	0.10 	&	$-0.76$ &0.36  	\\
ESZ-XMM-regular	&	0.50        	&	0.01                	&	0.98	&	0.10 	&	$-0.40$ &0.42  	\\
ESZ-XMM-irregular	&	0.49        	&	0.01                	&	1.00	&	0.10 	&	$-0.79$ &0.36  	\\
virial                	&	 \multicolumn{2}{c}{1/2}	         	&	 \multicolumn{2}{c}{1}               	&	 \multicolumn{2}{c}{0}   	\\
The 300        &	0.38       	&	0.01                	&	0.93	&	0.03 	&	$-0.12$ &0.05  	\\
\noalign{\smallskip}    
$E_\text{k}$-$E_\phi$ & \multicolumn{6}{c}{}	\\
\noalign{\smallskip}    
ESZ-XMM         	&	0.45       	&	0.03                	&	0.89	&	0.35 	&	$1.09$ &1.25  	\\
ESZ-XMM-regular	&	0.45        	&	0.05                	&	0.63	&	0.50 	&	$0.78$ &1.69  	\\
ESZ-XMM-irregular	&	0.45        	&	0.03                	&	0.89	&	0.36 	&	$1.05$ &1.23  	\\
virial                	&	 \multicolumn{2}{c}{1/2}	         	&	 \multicolumn{2}{c}{1}               	&	 \multicolumn{2}{c}{0}   	\\
The 300        &	0.43       	&	0.01                	&	1.00	&	0.03 	&	$-0.03$ &0.05  	\\
\noalign{\smallskip}   
	\hline
	\end{tabular}
}
\end{table}


\section{Results}

In our analysis, we consider the kinetic energy of galaxies. While cold and hot baryons interact, e.g. through galaxy motions driving turbulence, the kinetic energy of galaxies cannot be equated to the kinetic energy of the gas. Galaxies do not thermalise as they do not have well-defined temperatures. Thermalisation should strictly refer to the transformation of the kinetic energy of the gas to thermal energy. Here, we perform an indirect investigation, as the kinetic energy of the gas is not measured. Gas thermalisation is constrained by comparing the evolutions of dark matter, galaxies, and intracluster medium. In this context, the analysis of the $E_\text{k}$--$E_\phi$ relation constrains the degree of virial equilibrium of the cold baryons in galaxies and serves as a consistency check for the gas thermalisation inferred from the $E_\text{th}$--$E_\phi$ relation.

The fitting results, see Table~\ref{tab_scaling} and Fig.~\ref{fig_E_th_e_k_E_pot}, show a remarkable agreement with the self-similar prediction with one important difference. We find strong evidence for negative time evolution of the $E_\text{th}$--$E_\phi$ relation for the full sample, $p(\gamma<0)\sim98.3$ per cent. At a given potential energy, low redshift clusters are hotter than high-redshift ones \citep{lov+al20}.

The statistical accuracy on slope of the $E_\text{k}$--$E_\phi$ at a given redshift ($\delta \beta\sim 0.3$) is not enough to detect effects driven by baryonic processes, e.g. dynamical friction, merging with the central galaxy, tidal disruption, and accretion from the surrounding large-scale structure, which can make the relation steeper than the self-similar expectation, see Eq.~(\ref{eq_en_phi_en_k}).


\subsection{Thermalisation} 

We perform our analysis of thermalisation in terms of time evolution relative to the present-day era, when ESZ-XMM clusters are expected to be near equilibrium within $r_{500}$, see Sec.~\ref{sec_lowz}, rather than considering absolute values. Systematics effects, see Sec.~\ref{sec_syst}, can explain the off-set between simulated and observed clusters, see Fig.~\ref{fig_E_th_e_k_E_pot}. However, known systematics are not redshift dependent and should not impact the evolution analysis.

We can find the thermalisation epoch as the time when the ratio of thermal to potential energy is 95 per cent of the today value, see Fig.~\ref{fig_energy_ratio_z}. The threshold is met at $z_\text{th}=0.14\pm0.06$  (corresponding to $t_\text{lb}=1.8\pm0.7~\text{Gyr}$). This is a conservative criterion for relaxed clusters. Even accounting for non thermal pressure at the ten percent level at $z=0$, clusters deviating $\sim15$ per cent from the virial equilibrium are still considered relaxed \citep{cui+al18}.


\subsection{Regular clusters}

\begin{figure}
\resizebox{\hsize}{!}{\includegraphics{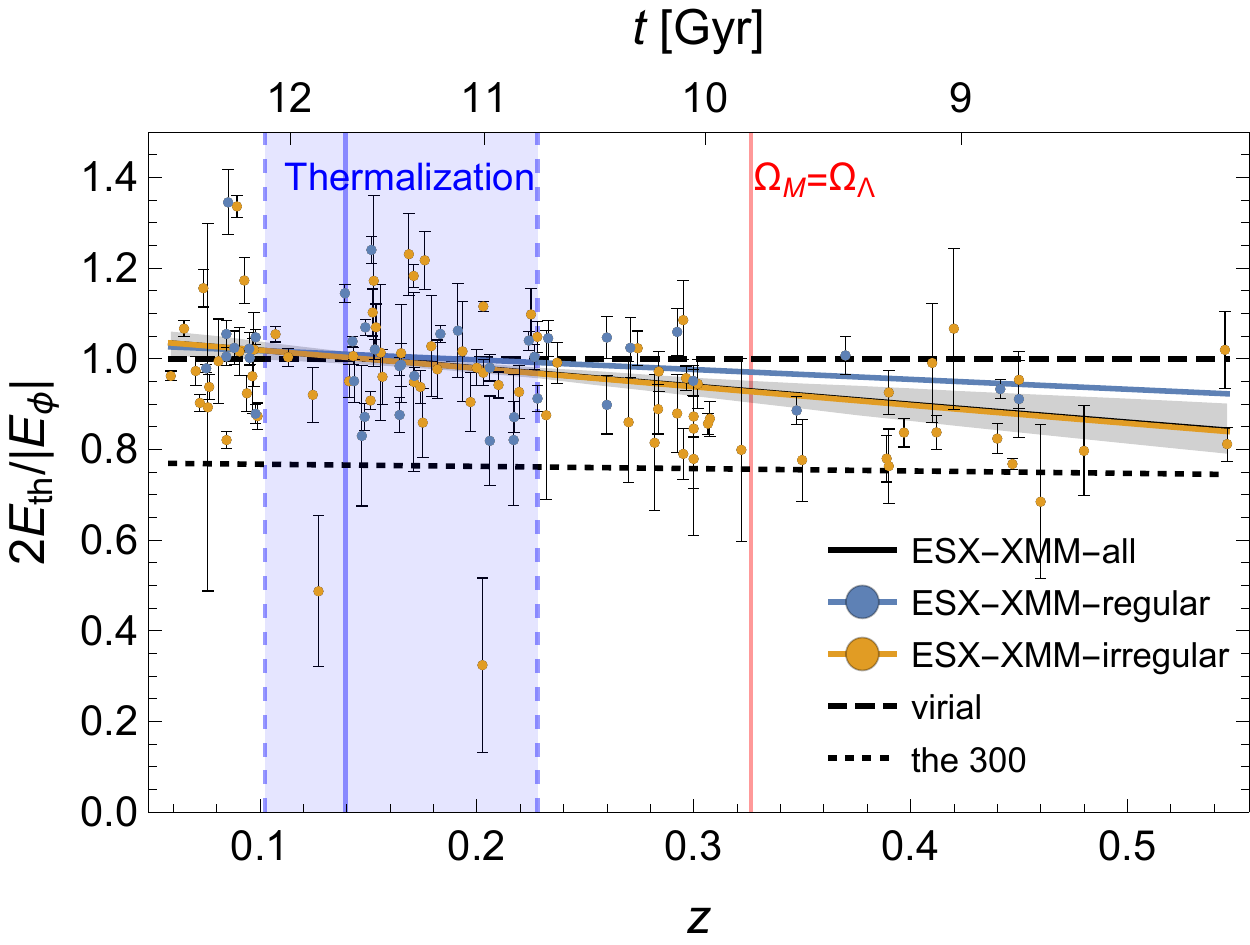}}
\caption{Time evolution. Shown is the time evolution of the ratio of twice the thermal energy $E_\text{th}$ to the absolute value of the potential energy $E_\phi$ for a cluster with $E_\phi \sim 18.7~\text{keV}$. The solid, dashed, or dotted lines show the fit to the ESZ-XMM clusters, the virial expectation, and the fit to the 300 simulations, respectively. Points and lines are colour-coded according to their morphological regularity with the 1/3 more regular clusters in cyan and the remaining 2/3 more irregular ones in orange. The shadowed region encompasses the 68.3 per cent confidence region around the fit for the full sample (solid black line). The vertical red and blue lines mark the dark matter -- dark energy equality and the thermalisation (with the associated 68.3 confidence region) time, respectively.
}
\label{fig_energy_ratio_z}
\end{figure}

The analysis of the regularity confirms our scenario. As expected, the negative evolution is driven mainly by irregular systems, whereas regular objects are compatible with no evolution, see Table~\ref{tab_scaling} and Fig.~\ref{fig_energy_ratio_z}. We find that the probability that the evolution of regular clusters is less pronounced (i.e. less negative) than that of irregular haloes is of $\sim 76.6$ per cent. On the other hand, the normalisation at $z=0$ is only slightly larger, which hints at low redshift ESZ clusters being generally relaxed. At a given time, differences in the dependence on the gravitational energy (as expressed by the slopes $\beta$) are not statistically significant.



\subsection{Simulations} 

We consistently use the definitions in Eqs.~(\ref{eq_en_phi}, \ref{eq_en_th}, \ref{eq_en_k}) for the analysis of either observations or simulations. For the simulated clusters, all measurements are performed in 3D, the core-excised temperature is mass weighted, and we consider the DM velocity dispersion. We consider the `true' temperature and we do not make use of {\it XMM}-like mocks of the 300.

The analysis of the 300 clusters supports a negative temporal evolution for the thermal energy, see Table~\ref{tab_scaling}. This confirms previous results which found a negative redshift evolution of the temperature with mass in simulated haloes \citep{leb+al17,tru+al18}. The slightly less than self-similar dependence of the thermal on the potential energy ($\Delta \beta \sim -0.07$, see Table~\ref{tab_scaling}) favours higher thermal content in less massive clusters, which are usually more relaxed and where feedback is more efficient. However, the statistical accuracy of our observations  ($\delta \beta \sim \pm0.10$) is not enough to significantly detect such variation. The scaling of the kinetic energy with the potential shows no significant deviations from self-similarity \citep{evr+al08}, which would be better appreciated at a lower mass range than what covered by ESZ \citep{mun+al13}.


\section{Systematics}
\label{sec_syst}

The energy ratios of systems in equilibrium might show fictitious time evolution due to systematics. The following effects are not included in our statistical model. They are expected to impact the normalisation of the scaling relations but not their slopes, which our conclusions are based on.

\subsection{Selection bias}
\label{sec_syst_sele}
The ESZ sample does not track an evolutionary sequence. The high redshift ESZ clusters at $z\sim 0.6$ are not the progenitors of the low redshift ones at $z\gtrsim 0$. Their descendants are more massive but we miss them due to limited observable volume of the local universe or selection effects. The regression analysis described in Sec.~\ref{sec_regr} can deal with observational biases and selection effects. The redshift dependent distribution of energies, see Eqs.~(\ref{eq_muz},~\ref{eq_siz}), can fit more massive systems at high redshift, see, e.g. \citet[][figures 5--12]{se+et15_comalit_IV} or \citet[][figures 5-7]{xxl_XXXVIII_ser+al19}.

However, we have still to assume that the not observed clusters share the same features of the observed sample. Our analysis looks for the epoch when a massive cluster of a fixed energy is in advanced equilibrium. The ESZ clusters are apt to this study since the median mass is sufficiently sampled over the redshift range, see Fig.~\ref{fig_ESZ_sample}, which mitigates the lack of massive local clusters in the sample.

\subsection{Temperatures}
\label{sec_syst_temp}

The measured spectroscopic temperature can be $\sim 10$ per cent larger than the mass-weighted value \citep{vik+al06}. ESZ-XMM benefits of an extended X-ray coverage by design, and we do not expect any redshift dependent temperature bias, even though uncertainties in the effective area calibration in the soft X-ray band could affect high and low redshift clusters differently \citep{lov+al20,lov+al20_mass}. 

A single, global temperature measurement cannot picture the details of the multi-phase gas distribution. We consider the core-excised temperature, which removes most of the multi-phase contribution and where residual effects of inhomogeneities and clumping are secondary. If we repeat our analysis by considering core-included temperatures to compute the thermal energy, we still find negative evolution with $\gamma=-0.61\pm0.38$ for the $E_\text{th}$--$E_\phi$ relation, in full agreement with the core-excised analysis, which confirms the secondary role of these effects for our results.

Systematic calibration effects and instrument-dependent biases make accurate X-ray temperatures very challenging. {\it XMM-Newton} temperatures of very hot plasmas are systematically lower than {\it Chandra}'s \citep{sch+al15}. High redshift ESZ-XMM clusters are on average more massive and hotter than local ones,
see Fig.~\ref{fig_ESZ_sample}. If the {\it Chandra} calibration was right and {\it XMM-Newton} was wrong, the temperatures of the high redshift clusters would be more biased low than the local ones, mimicking a negative evolution.

\subsection{Velocity dispersions}

Galaxy velocity dispersion estimates depend on the sampled members. Selection effects, galaxy luminosity and type, projected separation from the cluster center, or interlopers can bias the estimate. However, a sample of $\sim30$ redshifts of luminous  galaxies can be representative of the full cluster member population \citep{sar+al13}. The median number of spectroscopic redshifts per cluster of our sample is well above this threshold. The mean interloper fraction and aperture effects are nearly constant in the redshift and mass range covered by ESZ-XMM \citep{sar+al13}. The ratio of kinetic to potential energy is very responsive to the mass accretion history but can be overestimated in the inner regions with respect to larger radii \citep{pow+al12}.  DM velocity dispersions for ESZ-XMM can be overestimated by $\sim 5$ per cent  \citep{sar+al13,xxl_XXIII_far+al18}, with no time evolution.

\subsection{Masses}

Deviations from the spherical, isothermal density profile have to be accounted for in the computation of the energy terms. Due to turbulence and non-thermal pressure, the hydrostatic mass can be biased low by $\sim20$ per cent \citep{se+et15_comalit_I}. Non-thermal pressure can be more significant early. If hydrostatic masses were more underestimated at high redshift, a positive time evolution of the $E_\text{th}$--$E_\phi$ relation would be entailed, in opposition to our findings. However, the interplay between biases in temperature, non-thermal pressure, mass and gas density models, and hydrostatic mass is not trivial. Recent analyses confirmed that the hydrostatic bias dependence on mass and redshift is negligible up to $z\sim 0.8$ \citep{gia+al21}. 

Weak lensing masses are regarded as reliable mass measurements but they are not available for the full ESZ-XMM sample. By cross-matching with LC$^2$ \citep[Literature Catalogs of weak Lensing Clusters,][]{ser15_comalit_III}, which are standardised and homogenised compilations of clusters and groups with weak lensing mass estimates, we find weak lensing masses for 62 ESZ-XMM clusters\footnote{Updated LC$^2$  catalogs are available at \url{http://pico.oabo.inaf.it/\textasciitilde sereno/CoMaLit/LC2/}}. By comparison, ESZ-XMM hydrostatic masses are found to be biased low by $26\pm6$ per cent \citep{lov+al20_mass}.  The analysis of the weak lensing subsample confirms the negative evolution of the $E_\text{th}-E_\phi$ relation, $\gamma=-0.1\pm0.9$, but evidence is marginal since just 4 clusters at $z>0.4$ have known weak lensing mass. Based on these considerations, we do not expect that biases in the hydrostatic mass measurements can mimic a negative evolution of the $E_\text{th}$--$E_\phi$ relation.

\subsection{Density profiles}

The density profiles of the most massive clusters are expected to be self-similar to a very high degree from very early times \citep{leb+al18,mos+al19}.
The time evolution of cluster concentration \citep{men+al14} or triaxiality \citep{bon+al15} is negligible in the redshift range $0<z<0.6$. Gas thermodynamic properties show a remarkable similarity outside of the cluster core up to $z\sim 2$ \citep{mcd+al17,mos+al19,ghi+al21}. This is confirmed for the ESZ-XMM sample too, where the logarithmic slope of the gas density profile at $r_{500}$ for the clusters at $z<0.2$ $(\ge0.2)$ is $-1.94\pm0.46$ $(-2.03^{+0.34}_{-0.90})$. We do not expect any systematic effect in the analysis of the hydrostatic equilibrium due to the evolution of the gas structure that could mimic a negative evolution.

\subsection{Cosmological model}

Our analysis assumes a standard $\Lambda$CDM model. Cosmological parameters enter the calculation of the critical density and $H(z)$. Apparent time evolution might be the result of a wrong cosmology. 

The critical density does not depend on the normalisation of the power spectrum. Furthermore, we are interested in evolution relative to the present and our results do not depend on $H_0$. This makes our analysis independent from two of the most debated cosmological parameters, see e.g. \citet{koz+al19} or \citet{kids_hey+al21}. Discrepant estimates from different analyses might suggest a failure of $\Lambda$CDM at redshifts much larger than what we cover here ($z \la 0.6$).

An error on the comic matter density of $\delta\Omega_M \sim 0.05$, which is much larger than the statistical uncertainty of $\delta\Omega_M \la 0.01$ at the reach of today experiments, see e.g., \citet{planck_2018_VI}, would bring an error on $E_z$ at $z=0.5$ of $\ga 3$ per cent, i.e., a systematic error on the time evolution of $\delta \gamma \ga 0.1$, which is smaller than the precision we reach ($\delta \gamma \sim 0.4$).

\subsection{Summary}

Whereas the above known systematics cannot significantly affect the redshift evolution, they can impact the normalisations of the energy scaling relations. The measured present-day energy ratios are fully compatible with the expectation for isothermal clusters, but this is due to systematic effects, which bias the thermal or kinetic to gravitational energy ratio by $\sim 25$ per cent, and to the convenient choice for the numerical constant in our definition of $E_\phi$. Considering the spectroscopic temperature bias and that X-ray masses and temperature estimates are partially correlated, we expect the ratio of thermal to potential energy to be overestimated by $\sim 25$ per cent, which is fully consistent with the off-set between the ESZ-XMM and the 300 results ($\sim 30$ per cent, see Table~\ref{tab_scaling}). The galaxy velocity dispersion bias is comparable to the spectroscopic temperature bias, which makes the observed kinetic-to-thermal energy ratio nearly unbiased.

Our results do not depend on the normalisations of the energy ratios. They depend only on their evolution with respect to the present-day values. The previous known systematics are not expected to be redshift dependent. We can conclude that potential systematics do not impact significantly our results and our determination of the thermalisation epoch based on the $E_\text{th}$--$E_\phi$ relation. This conclusion is further supported by our findings on the kinetic energy, since a systematic redshift dependent mass bias, which is the larger contributor to the systematic budget, would bring on a fictitious negative evolution in the $E_\text{k}$--$E_\phi$ relation too, which we do not observe.

This conclusion could change if unknown systematics were to play a significant role. Selection effects combined with not accounted for cluster evolution, see Sec.~\ref{sec_syst_sele}, or temperature dependent calibration failures, see Sec.~\ref{sec_syst_temp}, could potentially mimic a negative redshift evolution.



\section{Discussion}
\label{sec_disc}


\subsection{Low redshift ESZ-XMM halos}
\label{sec_lowz}

The low redshift clusters of our sample can be seen as relaxed in the inner regions. The dynamical time-scale for a present-day halo is $\tau_\text{dyn}\sim2.8$~Gyr, see App.~\ref{sec_tim_sca}. Our sample covers nearly two dynamical time-scales, of the order of the relaxation time-scale needed for clusters to adapt after major mergers \citep{nel+al14,che+al19}. Haloes with early formation times are likely relaxed \citep{pow+al12,mos+al19}. 

Mergers with mass ratios less than 1:10 have only minor impact on the overall dynamical state \citep{che+al19}. Clusters at the low redshift end of our sample have typically $M_{500}\sim4\times10^{14}M_\odot$ and $k_\text{B}T_\text{g}\sim 5~\text{keV}$. According to theoretical models for the halo mass accretion history in $\Lambda$CDM \citep{gio+al12c,cor+al15}, such a present-day ($z=0$) cluster has on average accreted $\sim90$ percent of its final virial mass by $z\sim 0.11$ ($t_\text{lb}\sim1.4~\text{Gyr}$). 

The inner region within $r_{500}$ is more advanced in its relaxation process than the full virial region. The sound-crossing time within $r_{500}$ for the median ESZ cluster at low redshift is $\tau_\text{sc}\sim 1.2$~Gyr, see App.~\ref{sec_tim_sca}. The inner gas of our present-day clusters has then thermalised in a regime of mainly smooth accretion which does not significantly alter the hydrostatic equilibrium.

This scenario is also supported by recent observations. The ICM in the core of the Perseus cluster is found to be quiescent, with low levels of turbulence \citep{hitomi16}. Deep X-ray coverage out to the virial radius of local, massive, {\it Planck} clusters shows density and pressure profiles consistent with an ideal gas in hydrostatic equilibrium within a DM potential well \citep{ghi+al19} and mild levels of non-thermal pressure \citep{eck+al19}.

We can then consider the core-excised region within $r_{500}$ of present-day ESZ clusters as nearly thermalised.


\subsection{Gravitational and radiative processes}

In the most massive observed clusters, the evolution of the thermal content within $r_{500}$ is due to gravitational interactions more than radiative physics \citep{leb+al17,tru+al18,ghi+al19}. Baryonic processes are most effective in small haloes. At a given epoch, we do not find any statistically significant deviation of kinetic or thermal energy from the self-similar dependence on the potential energy. This strengthens the view that radiative processes play a major role only in small groups \citep{lov+al15,ser+al20_hscxxl}, at very high redshift $z\sim1$--$2$ \citep{tru+al18} , or in the very inner core \citep{mcd+al17,sie+al18,ghi+al19,say+al21}, which we excise from the analysis of the thermal content. 

Bulk flows and turbulence can act as a source of non-thermal pressure support which is more substantial at early times, when the kinetic energy of the hot gas is not yet converted in thermal energy. At fixed total energy, a lower thermal energy (with respect to low-redshift clusters) is accumulated within the overall potential well at high redshift, when most massive clusters are still far from equilibrium. The observed negative evolution (with respect to self-similarity) of the $E_\text{th}$--$E_\phi$ relation shows that haloes of fixed energy have decreasing thermal-to-kinetic energy ratios with increasing redshift. 

This feature is gravitationally driven and nearly independent of gas physics.
The thermal properties in the inner regions beyond the cooling core are mainly determined by the merging history of the cluster and are less affected by the current accretion rate, which determines the state of the gas in the outskirts \citep{ghi+al19}. The properties of the gas are remarkably self-similar and mostly follow the predictions of simple gravitational collapse with minor non-gravitational corrections.


\subsection{Accretion and merger history}

Shock heating due to minor or major mergers is effective in thermalising massive clusters. The thermalisation epoch appears to be strictly connected to the mass accretion history. The clusters we observe at intermediate redshift evolve into more massive haloes. ESZ-XMM clusters at $z_\text{th}$ have typically $M_{500}\sim5.2\times10^{14}M_\odot$ and $k_\text{B}T\sim 5.5~\text{keV}$. At $z_\text{th}$, they have already accreted on average $\sim 80$--$90$ per cent of their final ($z=0$) virial mass \citep{cor+al15}. They face an evolution mainly characterised by gentle smooth accretion and minor mergers that cannot significantly perturb the equilibrium in the inner regions. They experienced a quite calm evolution in their nearest past too. During the time corresponding to one collision time for the inner $r_{500}$ before $z_\text{th}$ ($\tau_\text{sc}\sim 1.0~\text{Gyr}$), they accreted on average $\sim 10$ per cent of their virial mass in the dark matter dominated universe.

The negative evolution of the thermal-to-potential energy is a very peculiar feature. Self-similar, isolated, and steady haloes in virial and hydrostatic equilibrium do not show any evolution. In the alternative inside-out formation scenario, a fast growth phase is followed by secondary infall and slow accretion from the surrounding environment \citep{fuj+al18a}. Even though virial equilibrium is not assumed, the ratio of thermal-to-potential energy is still nearly constant with time, with a possible but small positive evolution \citep{fuj+al18b}. Negative evolution is a sign of merger events  and not gentle accretion which temporarily disrupt the hydrostatic equilibrium. 

In the $\Lambda$CDM bottom-up scenario of cosmic structure formation, perturbations grow quickly at early times and nearly stall in the dark energy dominated phase. Mergers that can disrupt the virial equilibrium are more likely at high redshift. According to numerical simulations \citep{fak+al10}, only $\sim 15$ per cent of clusters at $z_\text{th}$ with a gravitational energy of $E_\phi \sim 18.7~\text{keV}$, the median value for the ESZ-XMM sample, are experiencing a major merger with mass ratio of the progenitor haloes larger than 0.2.

Negative evolution of the thermal to potential energy ratio can naturally come out of an evolutionary sequence. Thermalisation operates by transferring kinetic to thermal energy. Dynamically not relaxed clusters should be in a relatively earlier stage of thermalisation than relaxed clusters as the kinetic energy during or just after mergers has still to settle. Relaxed clusters normally form early and their degree of equilibrium increases with time \citep{mos+al19}. Unrelaxed clusters form late. At a given redshift, the progenitor of a low redshift, unrelaxed cluster should be in a relatively earlier stage of thermalisation and with a lower thermal energy than the progenitor of a relaxed cluster, whose formation time is expected at higher redshift \citep{mos+al19}.


\subsection{Previous results}

The relation between thermal and potential energy is a convenient recasting of the mass--temperature relation, which has been a major topic of interest in cluster astrophysics \citep{gio+al13}. We refer to \citet{lov+al20} for a full discussion of the mass--temperature relation for the ESZ-XMM sample. \citet{lov+al20} considered one-on-one relations, whereas our analysis exploits a joint multi-fit between three observables which enable us to study the co-evolution of hot and cold baryons and dark matter.

A number of analyses of the mass-temperature relation consider either a self-similar time evolution \citep{vik+al09}, or local \citep{lov+al15,sch+rei17} or high redshift \citep{jee+al11,sch+al18} samples. Here we briefly discuss some previous results which investigated redshift evolution. \citet{kot+vik05} presented an X-ray analysis of 10 galaxy clusters at $0.4\la z \la 0.7$ with publicly available {\it XMM-Newton} observations spanning a temperature range $2.5~\textrm{keV} \la T \la 9~\textrm{keV}$. Based on hydrostatic masses, they found $\gamma_{E_\text{th} | E_\phi}\simeq -0.08\pm0.23$.

\citet{rei+al11} compiled an heterogeneous list of 232 clusters with $T \ga 2~\textrm{keV}$ out to $z\sim1.5$ from literature or new observations. Temperatures were based on {\it XMM } or {\it Chandra} observations and masses were measured exploiting different proxies. After homogenisation of the sample, they found $\gamma_{E_\text{th} | E_\phi}\simeq 0.02\pm0.07$.

\citet{man+al16_wtg} presented constraints on the scaling relations of galaxy cluster X-ray luminosity, temperature, and gas mass with mass and redshift. They analysed clusters from an X-ray flux-limited sample in the redshift range $0\la z\la 0.5$. Centre-excised X-ray temperatures from {\it Chandra} were available for 139 clusters and weak lensing masses for 27 clusters. They found $\gamma_{E_\text{th} | E_\phi}\simeq -0.08\pm0.20$.

\citet{bul+al19} analysed {\it XMM} X-ray observations of a SZ selected sample of 59 galaxy clusters from the South Pole Telescope SPT-SZ survey with masses $M_{500} \ga 3\times10^{14}M_\odot$ that span the redshift range $0.20\la z\la1.5$. Employing halo masses based on SZ observable-to-mass scaling relations and calibrated using information that includes the halo mass function, and core excised temperatures, they found $\gamma_{E_\text{th} | E_\phi}\simeq -0.57\pm0.25$.

\citet{ser+al20_hscxxl} exploited multi-wavelength surveys -- the XXL survey at {\it XMM-Newton} in the X-ray band, and the Hyper Suprime-Cam Subaru Strategic Program for optical weak lensing -- to study an X-ray selected, complete sample of clusters and groups. With a multivariate analysis of gas mass, core-included temperature, soft-band X-ray luminosity, and weak lensing mass of 97 groups with  $M_{500}\sim 1.3\times10^{14}M_\odot/h$ in the redshift range $0\la z\la 1$, they found $\gamma_{E_\text{th} | E_\phi}\sim -0.67\pm0.57$.

Notwithstanding the very different mass calibrations, sample selections, X-ray observatories or analyses, and statistical approaches, the above studies mostly point to negative evolution and hotter clusters at low redshift, in agreement with our findings. Negative evolution was detected with larger statistical evidence in analyses of SZ selected samples, e.g. \citet{bul+al19} and the present analysis. This might be due to SZ samples covering a diverse population, wheres X-ray selection could suffer from a cool core bias and favour more relaxed, centrally concentrated systems  \citep{ros+al17}, that near equilibrium at earlier epochs.

Even though the statistical evidence for each result is at the $\sim1$-2  $\sigma$ level, we notice that the probability of 6 coin flips (including our results) to give 0 or 1 heads (or tails) is of $\sim89$ per cent. However, we caution against over-interpretation due to the partial and incomplete review of previous results.


\subsection{Clusters as cosmological probes}

Probes of the thermal content of galaxy clusters, either based on X-ray observations of gas mass and temperature \citep{kra+al06} or SZ signal, are routinely used as mass proxies for scaling relations and cosmological constraints based on number counts. The tension between the value of the normalised matter density parameter obtained from cluster number counts, and the estimates based on {\it Planck} analysis of primary fluctuations in the cosmic microwave background is usually imputed to the absolute mass calibration \citep{lin+al14,planck_2015_XXIV}. {\it Planck} calibrated masses are biased low with respect to weak lensing masses, with a bias more pronounced for high redshift clusters \citep{se+et17_comalit_V}. The late thermalisation of galaxy clusters and the redshift evolution of the thermal-to-gravitational energy ratio could explain some discrepancies. Our finding is in agreement with the marginal detection of negative redshift evolution between SZ signal and mass for {\it Planck} clusters \citep{se+et15_comalit_IV}.


\section{Conclusions}

We studied some of the most massive clusters in the {\it Planck} observed universe out to $z\sim0.6$. SZ selected samples are regarded as reliable tracers of the halo mass function and cover a very diverse population. SZ samples are then suitable to studies of evolution since they are not severely biased towards more regular and relaxed systems.

At a given energy, the gas in the region outside the inner core and within nearly half of the virial radius in high redshift systems is colder than in local systems. The gas kinetic energy has still to convert to thermal energy. The galaxies reach equilibrium in the dark matter dominated halo more quickly than the gas. The evolution of the cluster thermal content is due to gravitational interactions more than radiative physics, which mostly affect the inner core we exclude from the analysis. 

The gas takes some time to reach equilibrium. The thermalisation epoch follows the cosmic dark matter - dark energy equality and lies in a gentler era of structure growth. The negative evolution of the thermal properties of the regions beyond the core are mainly determined by the merging history of the cluster and are less affected by smooth accretion or radiative processes. The thermalisation of the gas in terms of the merger and accretion history of galaxy clusters can be seen as another success of the $\Lambda$CDM paradigm.

Our results were based on a multi-variate analysis of the potential, kinetic, and thermal energy of the clusters. Analysis in terms of energy rather than mass can make the interpretation of results simpler. Energies are also simpler to measure than masses. Joint analysis of multiple cluster properties can be very helpful to better understand the formation and evolution of these systems \citep{ser+al17_CLUMP_M1206,ser+al18_CLUMP_I,say+al21}.

\section*{Acknowledgements}

The authors thank Elena Rasia for providing data on the 300 simulations, discussions, and a critical reading of the paper. MS thanks Carlo Giocoli and Stefano Ettori for fruitful discussions.
MS acknowledges financial contribution from contract ASI-INAF n.2017-14-H.0. LL and MS acknowledge financial contribution from contract INAF mainstream project 1.05.01.86.10.  LL acknowledges financial contribution from the contract ASI-INAF Athena 2019-27-HH.0, `Attivit\`a di Studio per la comunit\`a scientifica di Astrofisica delle Alte Energie e Fisica Astroparticellare' (Accordo Attuativo ASI-INAF n. 2017-14-H.0). WC acknowledges the support from the European Research Council under grant 670193. GS acknowledges support provided by the National Aeronautics and Space Administration (NASA) through Chandra award Number GO5-16126X issued by the Chandra X-ray Observatory Center (CXC), which is operated by the Smithsonian Astrophysical Observatory (SAO) for and on behalf of NASA under contract NAS8-03060.

\section*{Data availability}

The data underlying this article will be shared on reasonable request to the corresponding author.








\appendix

\section{Mass--temperature}
\label{sec_m-t}

Analyses in terms of energy can be convenient, see Eqs.~(\ref{eq_en_phi}, \ref{eq_en_th}, \ref{eq_en_k}). We can compare results in terms of mass-temperature to ours by noticing that $E_\phi \sim (M_{500} E_z)^{2/3}$, and $E_\text{th} \sim T_\textrm{g}$, see Eqs.~(\ref{eq_en_phi}, and \ref{eq_en_th}). A result in the form 
\beq
T \propto M^{\beta_{T|M}} E_z^{\gamma_{M|T}} \, ,
\eeq
can be converted to our notation, 
\beq
E_\text{th}  \propto E_\phi^{\beta_{E_\text{th} | E_\phi}}E_z^{\gamma_{E_\text{th} | E_\phi}} \, ,
\eeq
by taking 
\begin{eqnarray}
\beta_{E_\text{th} | E_\phi} & = & 3/2 \beta_{T|M} \, , \\
\gamma_{E_\text{th} | E_\phi} & = & \gamma_{M|T}- \beta_{M |T} \, . 
\end{eqnarray}

Self-consistent analyses of scaling relations can lead to self-similar evolution of scaling relations that is significantly weaker than is commonly assumed \citep{mau14}. For the self-similar case, $\beta_{E_\text{th} | E_\phi}=1$ and $\gamma_{E_\text{th} | E_\phi}=0$ or $ \beta_{M |T} = \gamma_{M|T}=2/3$. Apparent time evolution in the mass--temperature relation, i.e. $ \gamma_{M|T}\neq2/3$, can be due either to true not self-similar time evolution ($\gamma_{E_\text{th} | E_\phi}\neq 0$) or not self-similar mass dependence ($\gamma_{E_\text{th} | E_\phi} = 0$ but $\gamma_{M|T} = \beta_{M |T} \neq 2/3$).

\section{Time-scales}
\label{sec_tim_sca}


The dynamical time-scale for a SIS does not depend on mass,
\begin{equation}
\tau_\text{dyn} \sim \frac{r_\Delta}{\sigma_\text{1D}} =\frac{2}{H(z)\sqrt{\Delta}} \sim 2.8~\text{Gyr} \left(\frac{70~\text{km~s}^{-1}\text{Mpc}^{-1}}{H(z)}\right) \left( \frac{100}{\Delta} \right)^{1/2} .
\end{equation}
For the relaxation time of the dark matter halo, we consider $\Delta_\text{vir}$, the virial overdensity \citep{br+no98}.

For the gas, we consider the time required for a sound wave to cross the region to be thermalised. Expressing the time-scale in terms of the gas temperature,
\begin{equation}
\tau_\text{sc} \sim \frac{r_\Delta}{\sigma_\text{1D}} \sim 1.1~\text{Gyr} \left(\frac{r_\Delta}{1~\text{Mpc}}   \right) \left( \frac{5~\text{keV}}{k_\text{B}T_\text{g}} \right)^{1/2} .
\end{equation}
Here, we are mainly interested in the thermalisation of the region within $r_{500}$.

We consider semi-analytical approximations for the mass accretion history \citep{cor+al15}.


\bsp	
\label{lastpage}
\end{document}